\documentclass[fleqn, usenatbib]{mnras}

\usepackage{newtxtext,newtxmath}
\usepackage[T1]{fontenc}
\usepackage{ae,aecompl}

\usepackage{graphicx}	        
\usepackage{amsmath}	        
\usepackage{url}
\usepackage{amssymb}	        
\usepackage{microtype}
\usepackage{xspace}
\usepackage{float}
\usepackage[usenames]{color}

\definecolor{green}{rgb}{0.3,0.7,0.}
\definecolor{orange}{rgb}{0.726, 0.015, 0.015}
\definecolor{skyblue}{rgb}{0, 0, 0.74}

\newcommand{\mini}{\ensuremath{M_{\rm ini}}\xspace}
\newcommand{\mfin}{\ensuremath{M_{\rm fin}}\xspace}
\newcommand{\lbol}{\ensuremath{L_{\rm bol}}\xspace}

\newcommand{\mcore}{\ensuremath{M_{\rm core}}\xspace}
\newcommand{\menv}{\ensuremath{M_{\rm env}}\xspace}

\newcommand{\yc}{\ensuremath{Y_{\rm c}}\xspace}
\newcommand{\msun}{\ensuremath{\mathrm{M}_{\odot}}\xspace}
\newcommand{\aov}{\ensuremath{\alpha_{\rm ov}}\xspace}
\newcommand{\teff}{\ensuremath{T_{\rm eff}}\xspace}
\newcommand{\amlt}{\ensuremath{\alpha_{\rm mlt}}\xspace}

\newcommand{\lsun}{\ensuremath{\mathrm{L}_{\odot}}\xspace}

\newcommand{\logg}{\ensuremath{\log g}\xspace}

\newcommand{\fovcore}{\ensuremath{f_{\rm ov,\,core}}\xspace}
\newcommand{\fovenv}{\ensuremath{f_{\rm ov,\,env}}\xspace}
\newcommand{\fovshell}{\ensuremath{f_{\rm ov,\,shell}}\xspace}

\newcommand{\lum}{\ensuremath{L}\xspace}
\newcommand{\mhecore}{\ensuremath{M_{\operatorname{He-core}}}\xspace}
\newcommand{\mesa}{\textsc{mesa}\xspace}
\newcommand{\snap}{\textsc{snapshot}\xspace}

\title[Uncertain Masses of SN Progenitors]{The Uncertain Masses of Progenitors of Core Collapse Supernovae and Direct Collapse Black Holes}

\author[E. Farrell et al.]{
Eoin J. Farrell$^{1}$\thanks{E-mail: efarrel4@tcd.ie},
Jose H. Groh$^{1}$,
Georges Meynet$^{2}$,
J.J. Eldridge$^{3}$
\\
$^{1}$School of Physics, Trinity College Dublin, The University of Dublin, Dublin, Ireland\\
$^{2}$Geneva Observatory, University of Geneva, Chemin des Maillettes 51, 1290 Sauverny, Switzerland\\
$^{3}$Department of Physics, Private Bag 92019, University of Auckland, New Zealand
}

\date{Accepted XXX. Received YYY; in original form ZZZ}

\pubyear{2019}

\begin{document}
\label{firstpage}
\pagerange{\pageref{firstpage}--\pageref{lastpage}}
\maketitle

\begin{abstract}
We show that it is not possible to determine the final mass \mfin of a red supergiant (RSG) at the pre-supernova (SN) stage from its luminosity \lum and effective temperature \teff alone. Using a grid of stellar models, we demonstrate that for a given value of \lum and \teff, a RSG can have a range of \mfin as wide as 3 to $45~\msun$. While the probability distribution within these limits is not flat, any individual determination of \mfin for a RSG will be degenerate. This makes it difficult to determine its evolutionary history and to map \mfin to an initial mass. Single stars produce a narrower range that is difficult to accurately determine without making strong assumptions about mass loss, convection, and rotation. Binaries would produce a wider range of RSG \mfin. However, the final Helium core mass \mhecore is well determined by the final luminosity and we find $\log (\mhecore/\msun) = 0.659 \log (L/\mathrm{L}_{\odot}) -2.630$. Using this relationship, we derive \mhecore for directly imaged SN progenitors and one failed SN candidate. The value of \mfin for stripped star progenitors of SNe IIb is better constrained by \lum and \teff due to the dependence of \teff on the envelope mass \menv for $\menv \lesssim 1~\msun$. Given the initial mass function, our results apply to the majority of progenitors of core collapse SNe, failed SNe and direct collapse black holes.
\end{abstract}


\begin{keywords}
stars: evolution -- stars: massive -- stars: supernovae
\end{keywords}

\section{Introduction}
Core Collapse Supernovae (CCSNe) have significant impacts across many areas of astrophysics, including chemical enrichment of the interstellar medium and galaxies, triggering of star formation and release of energy into their surroundings. Determining which stars explode and their final properties such as mass and chemical composition is an important open question in astrophysics.

One of the exciting advancements in the last two decades is the direct imaging of CCSNe progenitors in pre-explosion archival images (see reviews from \citealt{Smartt:2015}, \citealt{VanDyk:2017}, and references therein). The analysis of these observations, in combination with other techniques such as SN light curve modelling, can help us to make connections between CCSNe and their progenitor stars and to improve our understanding of the complexities and uncertainties in the evolution of massive stars. The fact that we know the evolutionary stage of CCSNe progenitors (i.e. they are the end stages of their lives) makes them especially useful for comparisons with stellar models.

\begin{figure*}
	\centering
	\includegraphics[width=0.64\hsize]{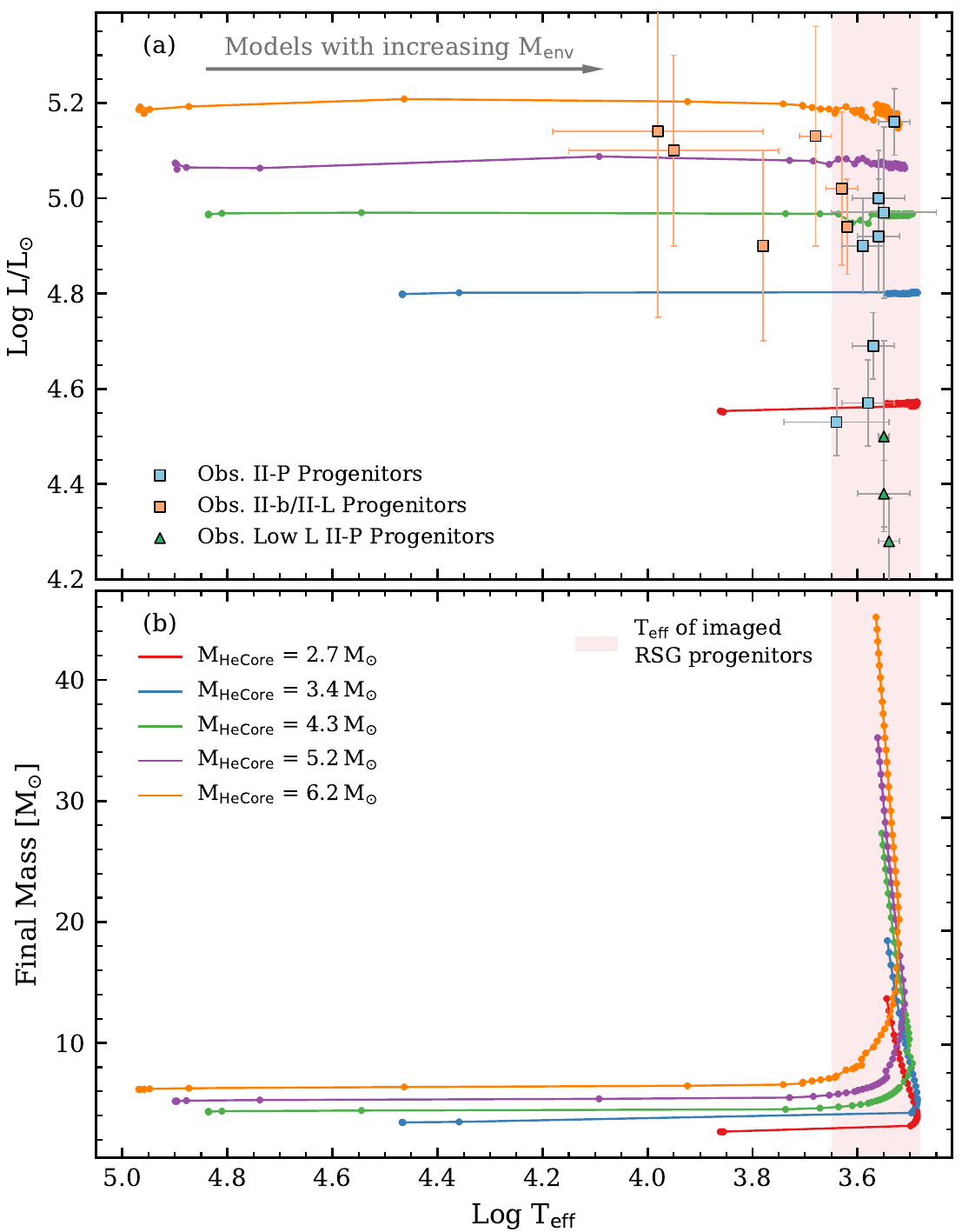}
	\caption{\textit{(a)}: Our grid of models at the end of central Carbon burning  in the HR diagram. Lines join models of the same Helium core mass and varying envelope mass from 0 to $\gtrsim 20\,\msun$. We also plot the \lum and \teff derived from pre-explosion images of progenitors of SN II-P (blue squares), IIb/II-L (orange squares) and low luminosity II-P (green triangles). See Table \ref{table:massfits} for further details. The Helium core masses are indicated in the legend in \textit{(b)}. We shade the range of \teff with which most RSG progenitors are observed (light red). \textit{(b)}: The final mass \mfin as a function of \teff for models with constant \mhecore and varying \menv.  \textit{(a)}.}
	\label{fig:presn_hrd}
\end{figure*}

The majority of CCSNe come from red supergiants (RSGs) with H-rich envelopes which explode as type-IIP SNe \citep[e.g.][]{Smartt:2009a, Smith:2011, Eldridge:2013, Groh:2013b}. Around 17 progenitors of CCSNe have been detected in pre-explosion images, the majority of which are RSGs. From these photometric observations, and with a distance, it is possible to obtain the bolometric luminosity \lbol and effective temperature \teff immediately before core collapse. To obtain an initial mass \mini from \lbol and \teff, it is necessary to use a stellar evolution model. Comparisons between the pre-explosion images and stellar evolution models have suggested that stars with $\mini \gtrsim 18~\msun$ may not explode as supernovae \citep[e.g.][]{Smartt:2009b}. However, some stellar evolution models predict that some stars with $\mini \gtrsim 18~\msun$ will die as RSGs. This discrepancy has been called the `red supergiant problem'. To explain this, several authors have proposed that RSGs with \mini between 18 and $30~\msun$ may collapse directly to a black hole without a luminous supernova explosion \citep[e.g.][]{Smartt:2015, Sukhbold:2016, Sukhbold:2019}. Others have offered suggestions related to underestimated bolometric corrections \citep{Davies:2018}, uncertain extinction \citep{Walmswell:2012}, and increased mass loss of luminous RSGs near the Eddington limit compared to older stellar models \citep[e.g.][]{Groh:2013b, Meynet:2015}. Note the statistical significance of the RSG problem is far lower than the original claim if late-type bolometric corrections are used \citep{Davies:2018}.

Uncertainties in stellar evolution models related to physical processes such as mass loss, convection, rotation and binary interaction mean that it is difficult to make a robust connection between observed surface properties of a progenitor and \mini. For instance, \citet{Groh:2013b} found that changes in the initial rotational velocity alone can cause an uncertainty of $\pm 2~\msun$ in the determination of the 
We should also keep in mind that it is possible that a significant fraction of RSG progenitors will have gained mass from a binary companion \citep[e.g.][]{Zapartas:2019}. This would produce RSGs with different core to envelope mass ratios than in single stars.

In this Letter we show that using the values of \lbol and \teff alone, it is difficult to derive the final mass of progenitors of SNe and direct collapse black holes. We also discuss the implications for their initial masses.

\vspace{-10pt}
\section{Stellar Models}
We compute a grid of stellar models at the end of core C burning spanning a range of He core masses \mhecore and envelope masses \menv. Our models have $\mhecore = 2.7$, 3.4, 4.3, 5.2 and $6.2~\msun$. For each \mhecore our grid contains models with \menv ranging from 0 to $\sim 40~\msun$. The envelopes of our models consists of $\sim 72$ per cent H in mass, except for models with $\menv \lesssim 0.5~\msun$ where the composition is not homogeneous. We choose this range of masses because they correspond to the majority of the range of observed CCSNe progenitors.

Our method can be summarised as follows. We first compute a stellar evolution model with the \mesa software package \citep[r10398,][]{Paxton:2011, Paxton:2013, Paxton:2015, Paxton:2018} from the zero-age main sequence until near the end of core-He burning. For these evolutionary calculations, we use standard physical inputs similar to \citet{Choi:2016}, with a solar metallicity of Z = 0.02. We pause the models when the central He abundance is $\yc = 0.01$. We then use a technique that we developed, named \snap, which allow us to add or remove mass from the star without the star evolving. In effect, this allows us to systematically modify \menv without affecting \mhecore. After \menv is modified, we allow the models to relax to hydrostatic and thermal equilibrium. Finally, we resume the evolution of these models until central C depletion with mass-loss turned off. The values of luminosity \lum and \teff at this point are the same as at the pre-supernova stage, as the surface properties are not expected to significantly change after central C depletion \citep{Groh:2013b, Yoon:2017}. The final stellar models provide the interior profiles of the standard quantities, e.g. chemical abundances, temperature, density and energy generation, in addition to the surface properties such as \lum and \teff.

The results from our models are subject to a number of caveats, which only add to our main conclusion that the initial and final masses of SN II-P progenitors are uncertain. We use standard mixing length theory for convection with a mixing-length parameter of \amlt = 1.82. This treatment of convection may affect the value of the stellar radius, and hence \teff. Secondly, we use a time-dependent, diffusive convective core-overshooting parameter \citep{Herwig:2000, Paxton:2011}. We adopt the same overshooting parameters as in the \textsc{mist} models \citep{Choi:2016} with core overshooting of \fovcore = 0.016 (roughly equivalent to \aov = 0.2 in the step overshoot scheme), and \fovshell = \fovenv = 0.0174. This may change the masses of the inert He shell and the mass of the CO core, which could have an impact on the core mass luminosity relationship that we derive. The nuclear reaction rates may also affect the core mass luminosity relationship. For instance, there is some uncertainty in the rate of $^{12}C(\alpha,\gamma)^{16}O$ \citep[e.g.][]{deBoer:2017}, which may impact the fractions of C and O in the core and hence the relationship between the core mass and \lum.

\section{The Uncertain Masses of Supernova Progenitors}


\begin{figure}
	\centering
	\includegraphics[width=0.95\hsize]{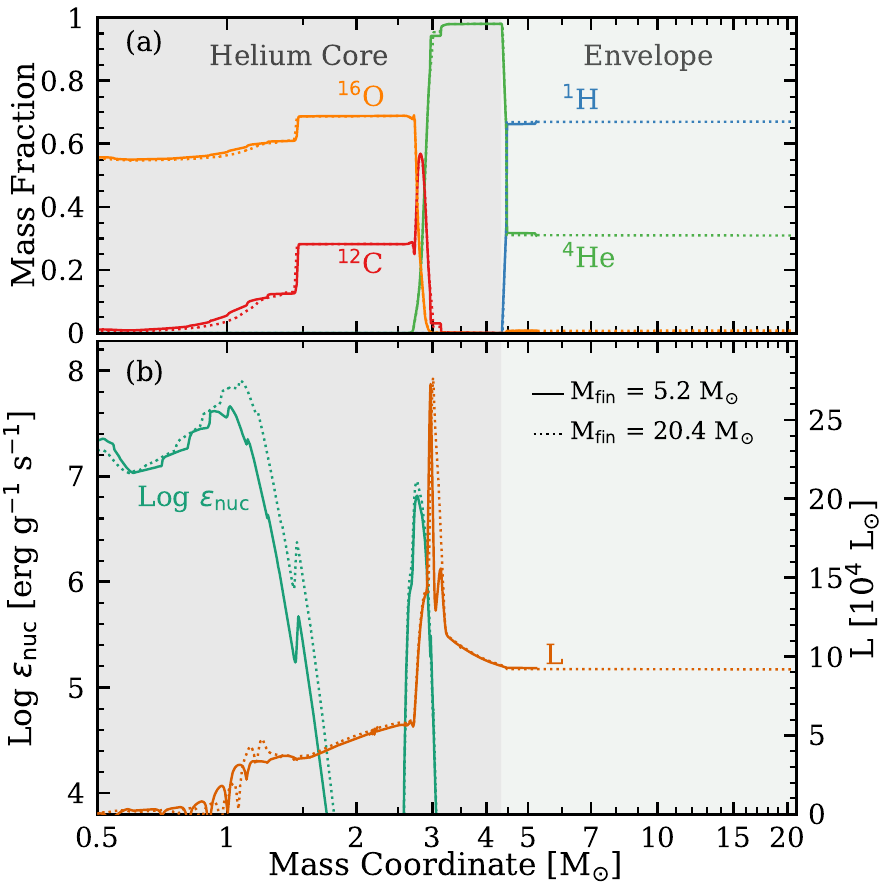}
	\caption{Comparing the interiors of two models at the end of central Carbon burning with the same Helium core mass, $\mhecore = 4.3~\msun$ and different final masses, $\mfin = 5.2$ (solid) and $20.4~\msun$ (dotted). We shade the Helium core in grey and the H rich envelope in light green.	\textit{(a)}: The internal abundance profiles of $^{1}$H (blue), $^{4}$He (green), $^{12}$C (red) and $^{16}$O (orange) as a function of Lagrangian mass coordinate in log scale. \textit{(b)}: The nuclear energy generation rate $\log \epsilon_{\rm nuc}$ in units of erg g$^{-1}$ s$^{-1}$ (green) and the internal luminosity profile \lum (orange).}
	\label{fig:internal_profiles}
\end{figure}

\begin{figure}
	\centering
	\includegraphics[width=0.98\hsize]{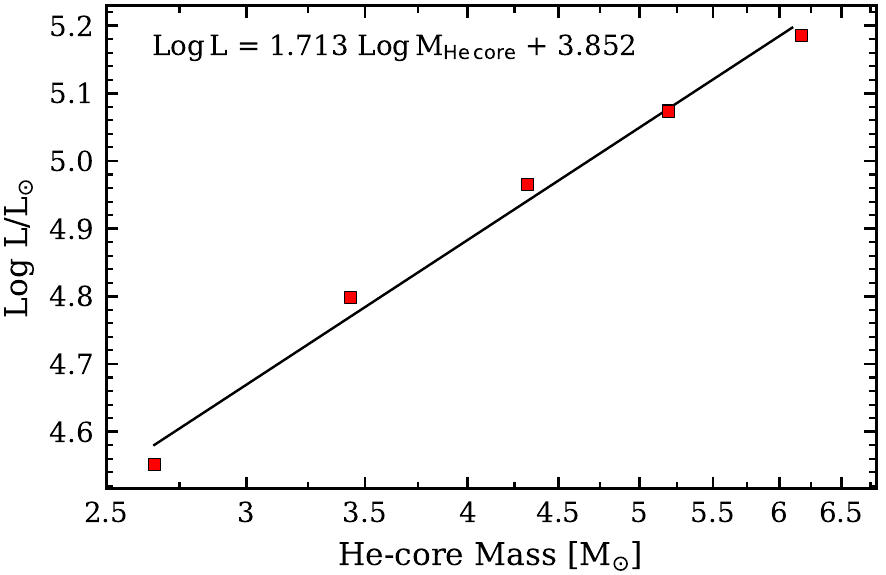}
	\caption{Relationship between final Helium Core Mass \mhecore and luminosity \lum of our models (red) at the end of central Carbon burning. We derive and plot the best fit relationship between \mhecore and \lum.}
	\label{fig:core_lum}
\end{figure}

\begin{table*}
\centering
\caption{Helium core masses \mhecore and range of allowed final masses \mfin that we derive from our models for a selection of progenitors type II-P, II-L and IIb supernovae as well as one direct collapse black hole candidate \citep[DCBH,][]{Adams:2017}. We take the values of \lum and \teff from 1. \citet{Aldering:1994}, 2. the compilation of \citet{Davies:2018}, 3. the updated distances for 2004et and N6946-BH1 provided by \citet{Eldridge:2019a}, 4. the compilation of \citet{Smartt:2015} and 5. \citet{Kilpatrick:2017}. We denote the value of \menv for progenitors of SN II-P by `...' as it cannot be constrained by \lum and \teff alone. We extrapolated our results to lower luminosities for the progenitors in italics. There is some debate about the progenitor of 2009kr \citep[See][]{Maund:2015}. We assume a minimum \menv of $1 \msun$ for progenitors of SN IIP.}
\label{table:massfits}
\begin{tabular}{llrrrrr r}
\hline
SN & Ref & Type & $\log\,(L/\lsun)$ & $\log T_{\rm eff}$ & \mcore [\msun] & \mfin [\msun] & \menv [\msun] \\
\hline
\textit{2003gd}  & 2 & II-P & 4.28 $\pm$ 0.09 & 3.54 & 1.8 $\pm$ 0.2 & 2.6 -- 13 & ... \\
\textit{2005cs}  & 2 & II-P & 4.38 $\pm$ 0.07 & 3.55 & 2.0 $\pm$ 0.2 & 2.9 -- 13 & ... \\
\textit{2009md}  & 2 & II-P & 4.50 $\pm$ 0.2 & 3.55 & 2.4 $\pm$ 0.7 & 2.8 -- 18 & ... \\
2008bk  & 2 & II-P & 4.53 $\pm$ 0.07 & 3.64 & 2.5 $\pm$ 0.2 & 3.3 -- 14 & ... \\
2012A  & 2 & II-P & 4.57 $\pm$ 0.09 & 3.58 & 2.6 $\pm$ 0.3 & 3.3 -- 14 & ... \\
2013ej  & 2 & II-P & 4.69 $\pm$ 0.07 & 3.57 & 3.1 $\pm$ 0.3 & 3.8 -- 18 & ... \\
2004A  & 2 & II-P & 4.90 $\pm$ 0.1 & 3.59 & 4.1 $\pm$ 0.6 & 4.6 -- 28 & ... \\
2012aw  & 2 & II-P & 4.92 $\pm$ 0.12 & 3.56 & 4.2 $\pm$ 0.7 & 4.6 -- 35 & ... \\
2006my  & 2 & II-P & 4.97 $\pm$ 0.18 & 3.55 & 4.5 $\pm$ 1.2 & 4.5 -- 45 & ... \\
2004et  & 3 & II-P & 5.00 $\pm$ 0.1 & 3.56 & 4.7 $\pm$ 0.7 & 5.1 -- 35 & ... \\
2012ec & 2 & II-P & 5.16 $\pm$ 0.07 & 3.53 & 5.8 $\pm$ 0.6 & 6.7 -- 41 & ... \\
N6946-BH1 & 3 & DCBH? & 5.50 $\pm$ 0.06 & 3.51 & 9.1 $\pm$ 0.8 & 9.4 -- 49 & ... \\
\hline
2011dh & 4 & IIb & 4.90 $\pm$ 0.2 & 3.78 & 4.1 $\pm$ 1.2 & 4.2$^{\scriptscriptstyle +1.3}_{\scriptscriptstyle -1.2}$ & 0.14$^{\scriptscriptstyle +0.13}_{\scriptscriptstyle -0.03}$ \\
2013df & 4 & IIb & 4.94 $\pm$ 0.1 & 3.62 & 4.3 $\pm$ 0.6 & 4.7$^{\scriptscriptstyle +0.8}_{\scriptscriptstyle -0.8}$ & 0.38$^{\scriptscriptstyle +0.25}_{\scriptscriptstyle -0.18}$ \\
1993J & 1 & IIb & 5.02 $\pm$ 0.16 & 3.63 & 4.8 $\pm$ 1.1 & 5.3$^{\scriptscriptstyle +2.3}_{\scriptscriptstyle -1.4}$ & 0.49$^{\scriptscriptstyle +1.18}_{\scriptscriptstyle -0.31}$ \\
2008ax & 4 & IIb & 5.10 $\pm$ 0.2 & 3.95 & 5.3 $\pm$ 1.6 & 5.5$^{\scriptscriptstyle +1.8}_{\scriptscriptstyle -1.7}$ & 0.22$^{\scriptscriptstyle +0.2}_{\scriptscriptstyle -0.09}$ \\
2009kr  & 2 & II-L & 5.13 $\pm$ 0.23 & 3.68 & 5.6 $\pm$ 2.0 & 6.1$^{\scriptscriptstyle +2.7}_{\scriptscriptstyle -2.3}$ & 0.49$^{\scriptscriptstyle +0.68}_{\scriptscriptstyle -0.32}$ \\
2016gkg & 5 & IIb & 5.14 $\pm$ 0.39 & 3.98 & 5.6 $\pm$ 3.8 & 5.8$^{\scriptscriptstyle +4.0}_{\scriptscriptstyle -3.9}$ & 0.23$^{\scriptscriptstyle +0.17}_{\scriptscriptstyle -0.11}$ \\
\hline

\end{tabular}

\end{table*}

Our models predict that it is not possible to determine the final mass, \mfin of a RSG supernova progenitor from \lum and \teff alone. For a given value of \lum and \teff, a RSG can have a range of \mfin as wide as 3 to $45~\msun$.

In Fig. \ref{fig:presn_hrd}a, we compare our grid of stellar models at the end of central C burning to the values of \lum and \teff derived from pre-explosion images of SN progenitors. The observations are taken from the compilation of \citet{Smartt:2015}. The models with high \teff furthest to the left in Fig.~\ref{fig:presn_hrd}a consist of a pure Helium core with no H-rich envelope. Moving from high to low \teff along each line corresponds to increasing \menv at constant \mhecore. For $\menv \lesssim 0.5 \msun$, the value of \teff decreases with increasing \menv due to the increased effect of opacity in the H-rich envelope. This effect has been seen before in single and binary stellar evolution models \citep[e.g.][]{Meynet:2015, Yoon:2017, Gotberg:2018}. For $\menv \gtrsim 0.5 \msun$, most models have a RSG structure with low \teff and a convective envelope and the value of \teff does not depend very strongly on the value of \menv. The value of \lum increases with \mhecore, however it does not depend on \menv (similar to the behaviour of \teff for RSGs). A given value of \lum can correspond to a wide range of \mfin. As a consequence of the relationship between the internal (\mhecore, \menv) and surface properties (\lum, \teff), there is a wide range of \mfin over which \lum and \teff are very similar.

To more clearly show the range of allowed masses for a given \lum and \teff, we plot the value of \mfin against \teff (Fig. \ref{fig:presn_hrd}b). As in Fig.~\ref{fig:presn_hrd}a, each line corresponds to a set of models with constant \mhecore and hence constant luminosity. For a given \mhecore, i.e. a given \lum, there is a large range of \mfin which produce similar values of \teff. Using our models, we estimate the range of allowed \mfin for a compilation of directly imaged SN progenitors (Table~\ref{table:massfits}).

To explore why the values of \lum and \teff of RSG progenitors are not strongly affected by \menv for a given \mhecore, we compare the interior of two models with the same final $\mhecore = 4.3~\msun$ and different final masses $\mfin = 5.2$ (solid) and $20.4~\msun$ (dashed) at the end of central Carbon burning (Fig. \ref{fig:internal_profiles}). 
The abundance profile of the core is very similar for both models (Fig.~\ref{fig:internal_profiles}a). The models with $\mfin = 5.2$ and $20.4~\msun$ have envelope masses of $\menv = 0.9$ and $16.2~\msun$ respectively. Fig.~\ref{fig:internal_profiles}b shows the nuclear energy generation rate  $\epsilon_{\rm nuc}$ (green) and the internal luminosity profile (orange) for each model. In both models, all of the nuclear energy generation occurs inside the Helium core. About 50 per cent of the luminosity from the core is generated by He-shell burning, above the CO core. No burning takes place in the H-shell. As a result, the internal luminosity profiles are very similar inside the core and constant outside the core. This results in the same surface luminosity for both models. The H-rich envelopes are fully convective in both models. This leads to a small change in radius, and hence \teff, over a wide range of \mfin.

While our models show that the values of \lum and \teff alone cannot determine \mfin, we can derive \mhecore from \lum (Table~\ref{table:massfits}). The dependence of the luminosity of RSGs on \mhecore has previously been pointed out \citep[e.g.][]{Smartt:2009a}. Knowledge of \mhecore is important for a number of reasons. For instance, the mass of the core determines the mass of the compact remnant left behind after the supernova, and also affects the nucleosynthesis and chemical yields. We derive the following relationship between the final core mass and the progenitor luminosity (Fig. \ref{fig:core_lum}):
\begin{equation} \label{eqn_l_to_mcore}
   \log (\mhecore/\msun) = 0.659 \log (L/\mathrm{L}_{\odot}) -2.630
\end{equation}
In terms of \mhecore, this is
\begin{equation} \label{eqn_mcore_to_l}
    \log (L/\mathrm{L}_{\odot}) = 1.713 \log (\mhecore/\msun) + 3.852
\end{equation}
The exponent in the core mass luminosity relationship of 1.713 is much lower than during core-He burning ($\sim 2.5$) or during the main sequence ($\sim 3.0$). It decreases as a massive star evolves.

For some of the progenitors in Table~\ref{table:massfits}, we have extrapolated Equation~\ref{eqn_l_to_mcore} to lower luminosities than we have modelled. We note that this makes those core masses very uncertain. The natures of the progenitors that have the lowest luminosities are uncertain \citep[e.g.][]{Eldridge:2007, Fraser:2011}. These stars are close to the minimum core mass for a core-collapse SN and expected to experience second dredge-up after core Helium burning and become AGB stars. If the low Helium core masses that we derive are correct and they do experience core collapse, it suggests that some physical process has slowed or prevented the process of second dredge-up. For example, \citet{Fraser:2011} found boosting the carbon-burning rate by a significant factor could prevent second dredge-up before core-collapse.  A detailed examination of whether models in this range would go through second dredge-up or not is beyond the scope of this work, but something we will investigate in future. For a review of the uncertain physics and outcomes see \citet{Doherty:2017}. We don't expect this to change the qualitative conclusion that the \mfin of RSG progenitors are uncertain.

We also use our models to derive \mhecore and \menv for 5 progenitors of SN IIb and II-L for which pre-explosion images exist (Table~\ref{table:massfits}). For models with $\menv \lesssim 1~\msun$, the value of \teff depends strongly on \menv. This allows a determination of \menv. The derived value of \menv depends strongly on both the values of \lum and \teff. The fact that \menv is well constrained means that the allowed rang of \mfin is much smaller for progenitors of SN IIb than for the RSG progenitors of SN II-P.

\section{Implications}
Our models predict that it is not possible to determine the mass of a RSG supernova progenitor from \lum and \teff alone. Based on the uncertainties in \lum and \teff, the range of allowed \mfin can be as wide as $3 - 45~\msun$ (Table~\ref{table:massfits}). While the probability distribution within these limits is not flat, and extreme values are unlikely, any determination of \mfin for a specific event based on the surface properties alone will be highly degenerate. RSGs that evolved through binary evolution can have a wider range of \mfin than single stars. This is particularly important if the binary fraction is high \citep{Zapartas:2019}. Additionally, \citet{Eldridge:2018} find that SN II-P like light curves can be produced from RSGs with $\mfin \sim 4\msun$, and that stellar mergers can produce RSGs with $\mfin \sim 40\msun$. For single stars, there is a much narrower expected range of final masses. However, accurate values are difficult to determine with current state-of-the-art stellar evolution models without making strong assumptions about mass loss, convection, and rotation.

While the value of \mfin is degenerate for a given \lum and \teff, it is possible to determine the value of \mhecore from \lum (Equation~\ref{eqn_l_to_mcore}). Using this, we derive values of \mhecore for a compilation of SN progenitors. We include uncertainties in the value of \mhecore based on the reported uncertainties in $L$. The apparent upper luminosity limit to RSG progenitors reported by \citet{Smartt:2015} of $\log L/\lsun \simeq 5.1 $~dex corresponds to a final \mhecore of $5.3 \msun$.
The distribution of final \mhecore may be a useful constraint for evolution models of massive stars. From the observational side, improvements in distance determination and reddening calculations can help to improve the accuracy of inferred final \mhecore. 

The mapping between the final \mhecore and the \mini depends on the uncertain physical inputs of the stellar evolution models such as mass loss, rotation, convection and binary interaction. This mapping is likely to be mostly affected by processes that modify the mass of the convective core during the main sequence (MS). The mass of the Helium core of a RSG progenitor is mostly determined at the end of the MS and not strongly affected by subsequent mass loss, binary interaction. Our results suggest that the `red supergiant problem' can be framed in terms of a mapping between \mini and final \mhecore. Uncertainty about the value of \mfin of RSG progenitors has several consequences. It means that a RSG progenitor with a given luminosity and \teff can be produced from a wide variety evolutionary histories. This makes it difficult to determine the lifetime of the star and to assign an age. This may be important to consider when assigning an age to a SN progenitor based on its mass and relating the age to the surrounding stellar population.

It is possible to break the degeneracy between L, \teff and \mfin of RSGs after they explode. One way is to use the light curve of the supernova to determine the mass of the H-envelope \citep[e.g.][]{Dessart:2019b}. The value of \menv can be added to the value of \mhecore derived from the luminosity of the progenitor to determine \mfin. It may also be possible to determine \mfin from the value of \logg, in the unlikely event that a spectrum of the progenitor is available. To make connections between \mfin and \mini, stellar evolution models are needed. For instance, by combining stellar evolution models of single and binary stars and explosion models, \citet{Eldridge:2019b} explored a wide range of light curve and progenitor properties of CCSNe.

In contrast to RSG progenitors of SNe IIP , the value of \mfin of stripped star progenitors of SN IIb/II-L is more well determined by the values of \lum and \teff due to the sharp dependence of \teff on \menv for $\menv \lesssim 1~\msun$ (Fig. \ref{fig:presn_hrd}b). The maximum \menv that we derive for progenitors of IIb is $0.49 \msun$. The range of allowed \mfin is mostly due to the uncertainty in \mhecore as a result of uncertain \lum. In addition, most of the uncertainty in the derived values of \menv is due to the uncertainty in the value of \lum. The derived values of \menv can help us to understand and provide useful constraints on stellar evolution, binary interaction and also be used as inputs to hydro-dynamic explosion models. Our models predict that for a star to be a RSG at the end of its evolution (assuming $\teff < 5000$~K), it must have \menv of $\gtrsim 0.1 - 0.5 \msun$, depending on the value of \mhecore. \citet{Eldridge:2018} found that the minimum hydrogen mass required to produce a SN II-P is $1 \msun$. RSGs with \menv of $\sim 0.1 - 1 \msun$ may produce SN II-L when they explode.

While the degeneracy between \mfin, \lum and \teff for progenitors of SN II-P can be broken using SN observations, this is obviously not possible for progenitors of failed supernovae such as N6946-BH1 reported by \citet{Adams:2017}.Assuming a RSG structure and the updated distance to its host galaxy reported in \citet{Eldridge:2019a}, we derive $\mhecore = 9.1 \pm 0.8$ and an allowed final mass range of $9 - 49 \msun$. This value is close to the Helium core mass for a black hole forming event assumed by \citet{Heger:2003} ($\sim 8 \msun$) and also by \citet{Sukhbold:2016}. Using the lower distance assumed in \citet{Adams:2017}, we derive $\mhecore = 7.2 \pm 0.6$. The core mass determines the outcome of stellar evolution and the lower and upper \mhecore for CCSNe will place tight constraints on stellar models. It is difficult to constrain the initial mass of a progenitor from its final \mhecore. There is no unique solution because of the multiple possible pathways to lead to the same final \mhecore.

For values of \menv higher than those depicted in Fig. \ref{fig:presn_hrd}b, our models produce blue supergiant (BSG) progenitors, similar to what has been seen in binary evolution models for mass gainers and mergers \citep[e.g.][]{Menon:2017}. In contrast to the RSG models, we find that the H-shell of BSG models is still generating energy at the end of central Carbon burning. This introduces additional complexities in deriving a relationship between \mhecore and \lum because there will be a contribution to \lum from the H-shell which will depend on \menv. In the future, we will compute a grid of BSG progenitor models at low metallicities  which has implications for the progenitor of SN1987A.

In this Letter, we discussed how the final masses of RSG progenitors of CCSNe, failed SNe and direct collapse black holes are difficult to derive from the luminosity and effective temperature alone. The mass of a RSG at the final stage of its evolution is very uncertain, regardless of the success of the explosion.


\bibliographystyle{mnras}
\bibliography{MyLibrary}

\label{lastpage}
\end{document}